\RequirePackage{fix-cm}
\documentclass[pdftex,twocolumn,epjc3]{svjour3}
\smartqed  
\usepackage{booktabs}

\usepackage{graphicx}
\graphicspath{{fig/}}
\usepackage{mathptmx} 
\usepackage[colorlinks, citecolor=blue, urlcolor=blue, linkcolor=blue, bookmarksdepth=3]{hyperref}
\usepackage{amsmath}
\DeclareMathOperator\erfc{erfc}
\usepackage{amssymb}
\usepackage{physics}
\usepackage{bm}
\usepackage[numbers,sort&compress]{natbib}

\usepackage{siunitx}

\journalname{Eur. Phys. J. C}
\begin{document}

\title{Prospect of undoped inorganic scintillators at 77 Kelvin for the detection of non-standard neutrino interactions at the Spallation Neutron Source}

\titlerunning{Undoped NaI/CsI at 77 K for NSI detection at the SNS} 

\author{Keyu Ding$^1$ \and Daniel Pershey$^2$ \and Dmitry Chernyak$^1$ \and Jing Liu$^{1,}$\thanksref{e2}}

\thankstext{e2}{e-mail: \href{mailto:jing.liu@usd.edu}{jing.liu@usd.edu} (corresponding author)}
\institute{Department of Physics, University of South Dakota, 414 East Clark Street, Vermillion, SD 57069, USA \and Department of Physics, Duke University, Physics Bldg., Science Dr.,Durham, NC 27708, USA}

\date{Received: date / Accepted: date}

\maketitle

\begin{abstract}
  Investigated in this work are sensitivities to non-standard neutrino interactions (NSI) of a prototype detector placed about 20 meters away from the Spallation Neutron Source at the Oak Ridge National Laboratory in two years of data taking. The presumed prototype consists of 10~kg undoped CsI scintillation crystals directly coupled with SiPM arrays operated at 77~K. Compared to the COHERENT CsI(Na) detector, a much higher light yield is assumed for the prototype. An experiment with a cylindrical undoped CsI crystal coupled directly to a photomultiplier tube at about 77~K was conducted to verify the light yield assumption.  A yield of $33.5 \pm 0.7$~photoelectrons per keV electron-equivalent (PE/keVee) was achieved in [13, 60]~keVee, which is much closer to the relevant energy region for the NSI search than some of the early studies.
\end{abstract}

\section{Introduction}
Forty three years after D. Freedman predicted the existence of coherent elastic neutrino-nucleus scattering (CEvNS)~\cite{freedman74}, it was experimentally observed by the COHERENT collaboration at the Spallation Neutron Source (SNS), Oak Ridge National Laboratory (ORNL), in 2017~\cite{coherent17}. The result has triggered lots of interest, not only because it confirmed a long predicted standard interaction that is important in the evolution of astronomical objects~\cite{janka07}, but also, and more importantly, because it demonstrates the possibility to probe a broad range of standard and new physics through the detection of low energy neutrino interactions, including nuclear form factors~\cite{patton12, patton13}, weak mixing angle~\cite{canas18} at low energies, neutrino electromagnetic interactions~\cite{kosmas15, gemma10, texono07, munu05}, sterile neutrinos~\cite{formaggio12}, and non-standard neutrino interactions (NSIs)~\cite{davidson03, bar05, coloma16, coloma17, liao17, papoulias18, denton18, nsi19}, etc.

NSIs, first mentioned by Wolfenstein in his paper introducing the matter effect on neutrino oscillations in 1978~\cite{wolfenstein78}, can be categorized into two types: neutral-current (NC) and charge-current (CC) NSIs. The Lagrangian of the former can be expressed as~\cite{coloma17}
\begin{equation}\label{e:nc}
  \mathcal{L}_\text{NC} = - 2 \sqrt{2} G_F \sum\limits_{f,P,\alpha,\beta} \varepsilon^{f,P}_{\alpha\beta}
  (\bar{\nu}_\alpha \gamma^\mu P_L \nu_\beta) (\bar{f} \gamma_\mu P f),
\end{equation}
where $G_F$ is the Fermi constant, $f$ is one of the charged fermions in $\{e,u,d\}$, $\{\alpha, \beta\}$ are flavor indices, $P\in\{P_L,P_R\}$ are the chirality projection operators, which can be parameterized into vector, $V$, and axial, $A$, components of the interaction. The $\varepsilon$ terms quantify the strength of the new interaction, $G_X$, with respect to the Fermi constant,  $\varepsilon^{f,P}_{\alpha\beta} \sim \mathcal{O}(G_X/G_F)$.

CC NSIs affect in general the production and detection of neutrinos; NC NSIs affect the neutrino propagation in matter~\cite{nsi19}, the introduction of which into the standard $3\times3$ neutrino mass and mixing scheme can hence change the whole picture of neutrino oscillation phenomenology, such as causing degeneracies in the measurement of the solar mixing angle~\cite{miranda06}, in deriving the CP-violating phase $\delta_\text{CP}$~\cite{liao16, flores18}, mass hierarchy~\cite{deepthi17}, etc. at current and future long-baseline neutrino experiments, such as DUNE~\cite{liao16, coloma16, coloma17}.

Given such an importance of NSIs, however, oscillation experiments are not sensitive to the terms that involve no flavor changing (or non-universal terms), $\varepsilon^{f,P}_{\alpha\alpha}$, which can be constrained better by neutrino scattering experiments, such as COHERENT.  A sizable $\varepsilon^{f,P}_{\alpha\alpha}$ will cause a change of the number of CEvNS events. One can hence estimate its significance by comparing the observed number of CEvNS events to that predicted by the Standard Model (SM).

In addition to helping pin down NSI parameters that cannot be constrained by neutrino oscillation experiments, the stringent constraint on NSIs can also help with direct dark matter detection experiments~\cite{Schumann19}. As the sensitivities of those experiments improves, coherent scatterings of solar neutrinos in their targets become a serious background (the so-called \emph{neutrino floor})~\cite{Billard:2013qya}. The introduction of NSIs results in an additional source of uncertainty in determining the level of the floor~\cite{nsi19}. The reduction of this uncertainty would consequently improve the sensitivity of direct dark matter search experiments deep underground.

The SNS at ORNL provides the world's most intense pulsed source of neutrinos~\cite{coherent18} in an energy region of specific interest for particle and nuclear astrophysics as a by-product of neutrons.  Interactions of a proton beam in a mercury target produce $\pi^+$ and $\pi^-$ in addition to neutrons. These pions quickly stop inside the dense mercury target. Most of $\pi^-$ are absorbed. In contrast, the subsequent $\pi^+$ decay-at-rest (DAR) produces neutrinos of three flavors.  The COHERENT experiment~\cite{coherent18} is an ensemble of neutrino detectors located along the Neutrino Alley~\cite{coherent17, coherent18} about 20 meters away from the source. Data taken with a 14~kg CsI(Na) detector~\cite{coherent17} and a 24~kg (active) liquid argon detector~\cite{coherent20} by the COHERENT Collaboration have already placed strong constrains on NSIs~\cite{coherent17, coherent20}.

The sensitivity of the inorganic scintillator based detector can be improved by the increase of the target mass and the decrease of its energy threshold as more CEvNS events are expected at lower energies~\cite{freedman74, bar05, coherent17, coherent20}. Two largest limiting factors in reducing the energy threshold of the CsI(Na) detector are~\cite{coherent17}, first, the Cherenkov radiation from charged particles passing through the quartz window of the photomultiplier tube (PMT) directly coupled to the CsI(Na) crystal, and second, the afterglow of the crystal itself after some bright scintillation events.

The first limiting factor can be eliminated by replacing the PMT with silicon photomultiplier (SiPM) arrays, which do not have a quartz window. However, SiPMs operated at room temperature exhibit much higher dark count rates (DCR) than PMTs~\cite{sipm}.  In order to reduce the DCR of SiPMs, they need to be cooled~\cite{akiba09, catalanotti15, ost15, igarashi16, aalseth17, giovanetti17}, for example by liquid nitrogen (LN2). The cryogenic operation calls for undoped CsI/NaI instead of doped ones, since the former at 77~K have about twice higher light yields than the latter at 300~K~\cite{Bonanomi52, Hahn53a, Hahn53, Sciver56, Beghian58, Sciver60, Fontana68, West70, Fontana70, Emkey76, persyk80, Woody90, Williams90, Wear96, Amsler02, Moszynski03, Moszynski03a, Moszynski05, Moszynski09, Sibczynski10, Sibczynski12, mikhailik15, csi, ess19}. The authors measured the light yield of undoped CsI at 77~K~\cite{csi, csi20} and recently achieved a yield of $\sim$26 photoelectrons (PE) per keV electron-equivalent or keVee using a cryogenic PMT with a peak quantum efficiency (QE) of $\sim$27\%.  A light yield of 40$\sim$50 PE/keVee is achievable if PMTs are replaced by SiPMs with a peak photon detection efficiency (PDE) of 40$\sim$50\%, which are already available in the market.

However, the high yields were measured at an energy range from 662 to 2614~keVee~\cite{csi, csi20}, far away from the region that is relevant to the CEvNS detection. There is already evidence of the non-linear scintillation responses of undoped NaI~\cite{Moszynski09} and CsI~\cite{Moszynski05} crystals. A measurement of the light yield at a lower energy region is needed to verify the feasibility of using undoped NaI and CsI at 77~K for CEvNS and NSI detections.

Reported in this paper is such a measurement using an undoped CsI crystal at 77~K with an $^{241}$Am source down to 13~keVee. Based on the measured light yield, the potential of a $\sim 10$~kg undoped CsI prototype detector located $\sim 20$ meters away from the SNS for the detection of NSIs is investigated.

\section{Light yield of undoped CsI down to 13 keVee}
\subsection{Experimental setup}
\label{s:exp}
\begin{figure}\centering
  \includegraphics[width=\linewidth]{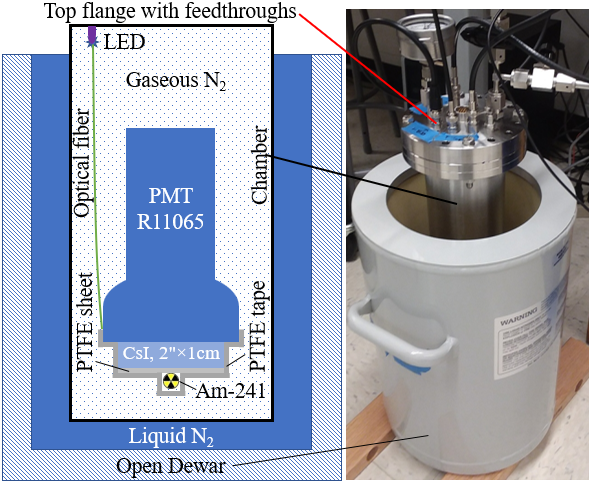}
   \includegraphics[width=\linewidth]{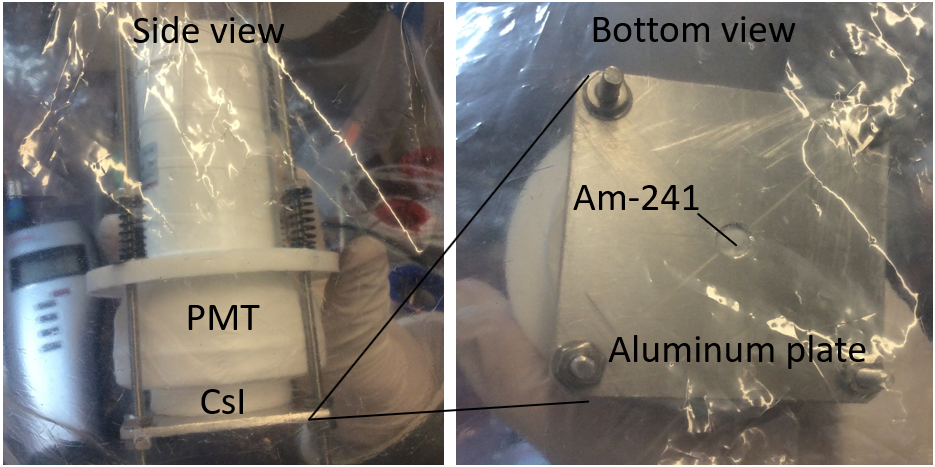}
  \caption{A sketch and pictures of the experimental setup.}
  \label{f:setup}
\end{figure}

Fig.~\ref{f:setup} shows the internal structure of the experimental setup for the measurement of the light yield of an undoped CsI crystal. The undoped cylindrical crystal was purchased from OKEN~\cite{oken}, and had a radius of 1~inch and a height of 1~cm. All surfaces were mirror polished. It was used in an earlier measurement, where a yield of $20.4\pm0.8$~PE/keVee was achieved above 662~keVee~\cite{csi}. Compared to the early measurement, the following modifications were made:

\begin{itemize}
  \item The side surface of the crystal was wrapped with multiple layers of Teflon tapes instead of a single layer to make sure that there was no light leak.
  \item The 2-inch Hamamatsu PMT R8778MODAY(AR) was replaced by a Hamamatsu 3-inch R11065-ASSY.
  \item In both setups, the PMTs were pushed against one of the crystal end surfaces by springs to ensure a good optical contact without optical grease. However, in the previous setup, the crystal was pushed against the bottom flange of the chamber, while in this setup, the crystal was pushed against an aluminum plate with a hole in the middle, leaving space for the placement of an $^{241}$Am source.
  \item The other end surface of the crystal was pushed against a PTFE sheet in between the crystal and the aluminum plate. The $^{241}$Am source was placed on the other side of the PTFE sheet so that alpha radiation was blocked from reaching the crystal.
\end{itemize}

To minimize exposure of the crystal to atmospheric moisture, the assembly was done in a glove bag flushed with dry nitrogen gas. The relative humidity was kept below 5\% at 22$^{\circ}$C during the assemble process.

The PMT-crystal assemble was lowered into a 50~cm long stainless steel chamber from its top opening. The inner diameter of the chamber was $\sim 10$~cm. The chamber was vacuum sealed on both ends by two 6-inch ConFlat (CF) flanges. The bottom flange was blank and attached to the chamber with a copper gasket in between. The top flange was attached to the chamber with a fluorocarbon CF gasket in between for multiple operations. Vacuum welded to the top flange were five BNC, two SHV, one 19-pin electronic feedthroughs and two 1/4-inch VCR connectors.

After all cables were fixed beneath it, the top flange was closed. The chamber was then pumped with a Pfeiffer Vacuum HiCube 80 Eco to $\sim 1\times {10}^{-4}$~mbar. Afterward, it was refilled with dry nitrogen gas to 0.19 MPa above the atmospheric pressure and placed inside an open LN2 dewar. The dewar was then filled with LN2 to cool the chamber and everything inside. After cooling, the chamber pressure was reduced to slightly above the atmospheric pressure.

A few Heraeus C~220 platinum resistance temperature sensors were used to monitor the cooling process. They were attached to the side surface of the crystal, the PMT, and the top flange to obtain the temperature profile of the long chamber. A Raspberry Pi 2 computer with custom software~\cite{cravis} was used to read out the sensors. The cooling process could be done within about 30 minutes. Most measurements, however, were done after about an hour of waiting to let the system reach thermal equilibrium. The temperature of the crystal during measurements was about 3~K higher than the LN2 temperature.

The PMT was powered by a CAEN N1470A high voltage power supply in a NIM crate. The signals were fed into a CAEN DT5751 waveform digitizer, which had a 1~GHz sampling rate, a 1~V dynamic range and a 10 bit resolution. Custom-developed software was used for data recording~\cite{daq}. The recorded binary data files were converted to CERN ROOT files for analysis~\cite{nice}.

\subsection{Single PE response}
\begin{figure}[htbp]\centering
  \includegraphics[width=0.55\linewidth]{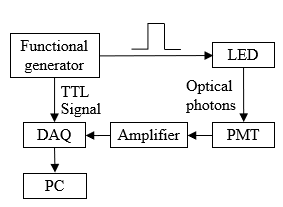}
  \caption{Trigger logics for the PMT single-PE response measurements.}
  \label{f:trg}
\end{figure}

The single-PE response of the PMT was measured using light pulses from an ultraviolet LED, LED370E from Thorlabs. Its output spectrum peaked at 375~nm with a width of 10~nm, which was within the 200 -- 650~nm spectral response range of the PMT. Light pulses with a $\sim$50~ns duration and a rate of 10~kHz were generated using an RIGOL DG1022 arbitrary function generator. The intensity of light pulses was tuned by varying the output voltage of the function generator so that only one or zero photon hit the PMT during the LED lit window most of the time. A TTL trigger signal was emitted from the function generator simultaneously together with each output pulse. It was used to trigger the digitizer to record the PMT response. The trigger logic flow chart is shown in Fig.~\ref{f:trg}.

The PMT was biased at 1,600~V, slightly above the recommended operation voltage, 1,500~V, to increase the gain of the PMT.  Single-PE pulses were further amplified by a factor of ten using a Phillips Scientific Quad Bipolar Amplifier Model 771 before being fed into the digitizer in order to separate them well from the pedestal noise.

\begin{figure}[htbp] \centering
  \includegraphics[width=\linewidth]{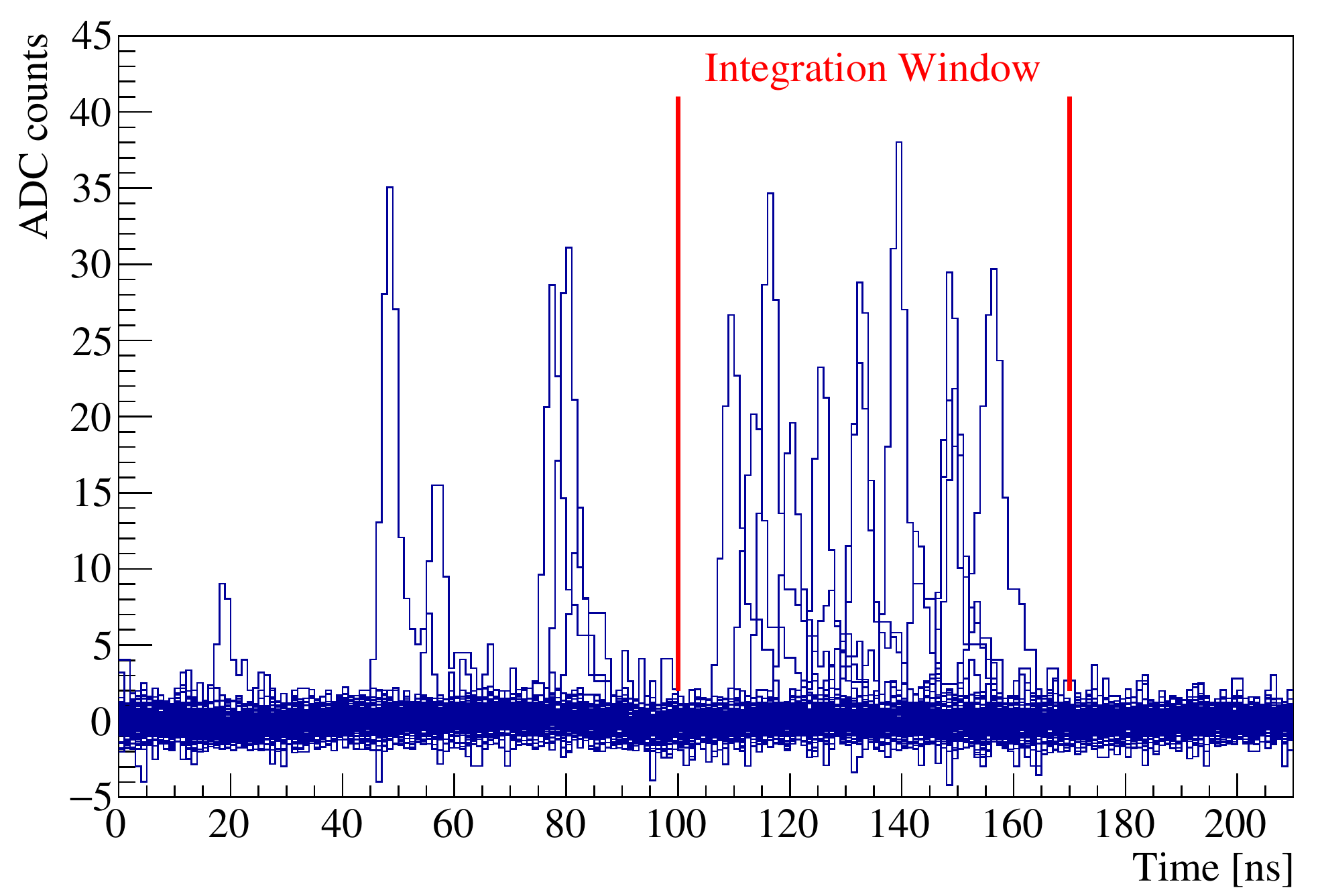}
  \caption{Two hundred consecutive waveforms from the PMT overlapped with each other measured with the crystal in place.}
  \label{f:ps}
\end{figure}

\begin{figure}[htbp] \centering
  \includegraphics[width=\linewidth]{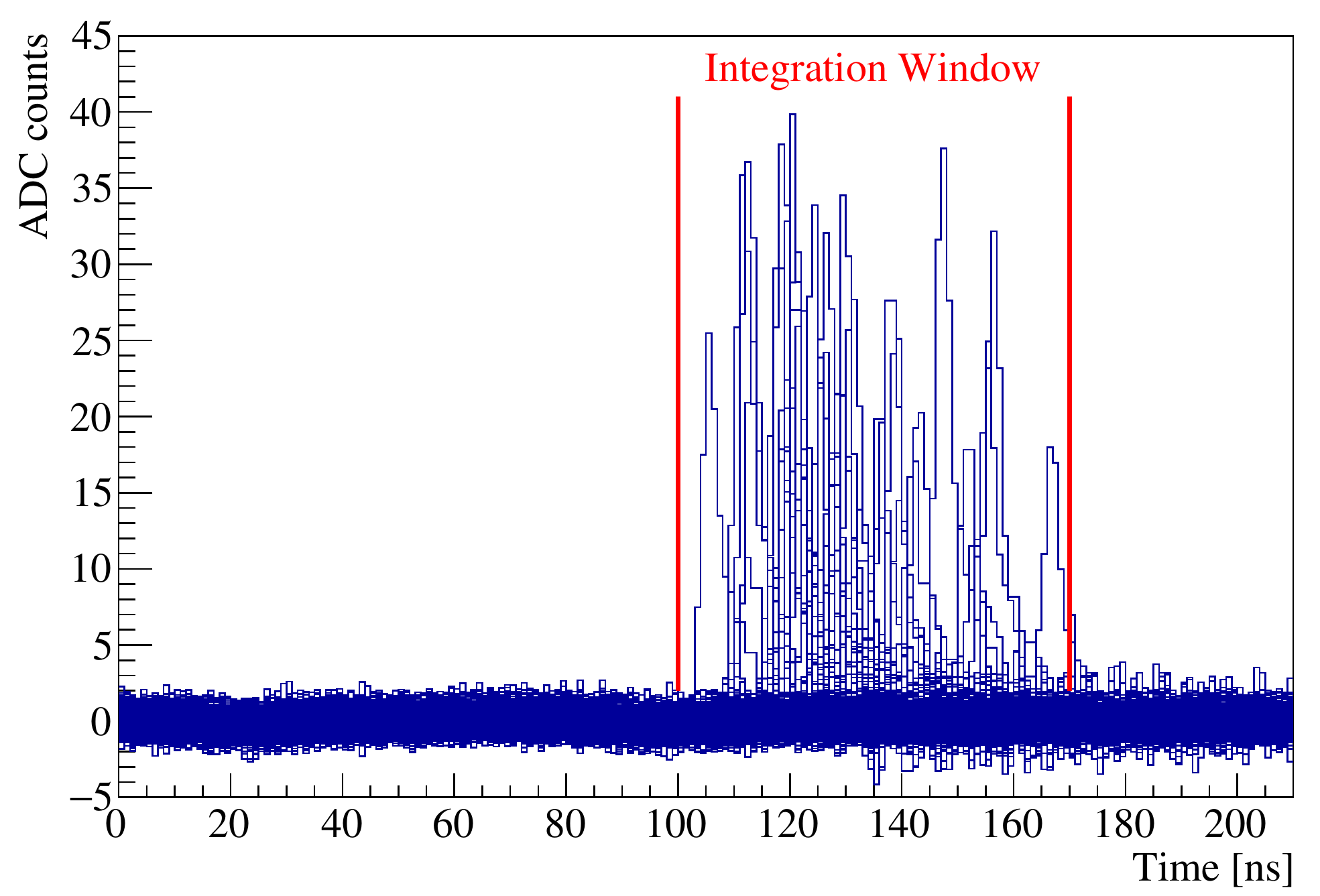}
  \caption{Two hundred consecutive waveforms from the PMT overlapped with each other measured without the presence of the crystal.}
  \label{f:psc}
\end{figure}

Fig.~\ref{f:ps} shows two hundred consecutive waveforms from the PMT randomly chosen from a data file taken during a single-PE response measurement. The integration window marked in the figure coincided with the LED lit window. Some single-PE pulses could be seen outside of the window. They were thought to be due to scintillation of random low energy radiation in the crystal.

To verify this assumption, the same measurement was repeated without the presence of the crystal. The resulting waveforms are shown in Fig.~\ref{f:psc}, where no pulse outside of the integration window can be seen.

\begin{figure}[htbp] \centering
  \includegraphics[width=\linewidth]{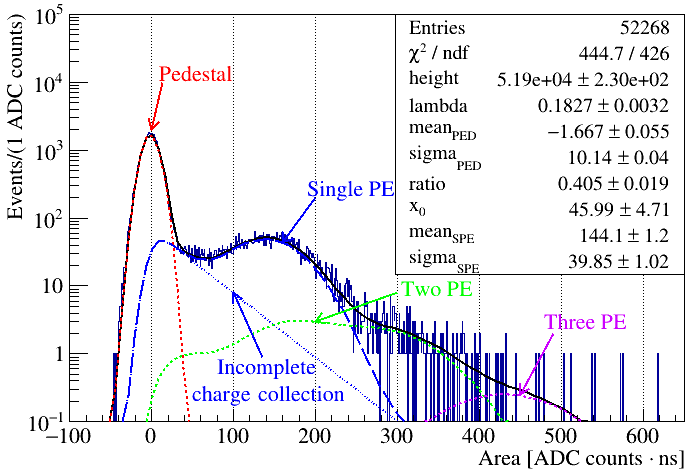}
  \caption{Single PE response of the PMT in logarithm scale.}
  \label{f:1pe}
\end{figure}

An integration in this time window was performed for each waveform in the data file whether it contained a pulse or not. The resulting single-PE spectrum is shown in Fig.~\ref{f:1pe}. The location of the single-PE peak varied within 5\% in different measurements. In the energy calibration measurement to be mentioned in a later section, the single-PE spectrum with the crystal was used but with a 5\% uncertainty attached to be conservative.

The spectrum was fitted in the same way as described in Ref.~\cite{ds1013} with a function,
\begin{equation}\label{e:Fx}
F(x)=H\sum\limits_n P(n,\lambda) f_n(x),
\end{equation}
where $H$ is a constant to match the fit function to the spectrum counting rate, $P(n,\lambda)$ is a Poisson distribution with a mean of $\lambda$, which represents the average number of PE in the time window, $f_n(x)$ represents the \textit{n}-PE response, and can be expressed as
\begin{equation}\label{e:fnx}
f_n(x)=f_0(x) \ast f_1^{n\ast}(x),
\end{equation}
where $f_0(x)$ is a Gaussian function representing the pedestal noise distribution, $\ast$ denotes a mathematical convolution of two functions, and $f_1^{n\ast}(x)$ is a n-fold
convolution of the PMT single-PE response function, $f_1(x)$, with itself. The single-PE response function $f_1(x)$ was modeled as:
\begin{equation}\label{e:f1x}
f_1(x)=
  \begin{cases}
    R(\frac{1}{x_0} e^{-x/x_0})+(1 - R)G(x;\bar{x},\sigma) & x>0; \\
    0 & x\leq0,
  \end{cases}
\end{equation}
where $R$ is the ratio between an exponential decay with a decay constant $x_0$, and a Gaussian distribution $G(x;\bar{x},\sigma)$ with a mean of $\bar{x}$ and a width of $\sigma$. The former corresponds to the incomplete dynode multiplication of secondary electrons in the PMT. The latter corresponds to the full charge collection in the PMT.

The fitting function has eight free parameters as shown in the top-right statistic box in Fig.~\ref{f:1pe}, where ``height'' corresponds to $H$ in Eq.~\ref{e:Fx}, ``lambda'' corresponds to $\lambda$ in Eq.~\ref{e:Fx}, ``mean'' and ``sigma'' with a subscript ``PED'' represents the mean and the sigma of the Gaussian pedestal noise distribution, those with a subscript ``SPE'' represents $\bar{x}$ and $\sigma$ in Eq.~\ref{e:f1x}, respectively, and ``ratio'' corresponds to $R$ in Eq.~\ref{e:f1x}.  Due to technical difficulties in realizing multiple function convolutions in the fitting ROOT script, the three-PE distribution, $f_1^{3\ast}(x)$, was approximated by a Gaussian function with its mean and variance three times that of the single-PE response.

Table~\ref{t:1PE} lists means of single-PE distributions measured before and after the energy calibration to be mentioned in the next section to check the stability of the PMT gain. The average mean for the PMT at 1,600~V is $14.5 \pm 0.1$~ADC counts$\cdot$ns after being divided by the amplification factor, 10.

\begin{table}[htbp] \centering
  \caption{\label{t:1PE} Summary of single-PE response measurements before and after the energy calibration to be mentioned in the next section.}
  \begin{tabular}{cccc}
    \toprule
     &Temperature &Temperature &Mean$_\text{SPE}$ \\
    &of PMT [$^\circ$C]&of crystal [$^\circ$C] &[ADC counts$\cdot$ns]\\
    \midrule
     Before & -193.8 $\pm$ 1.1 & -195.7 $\pm$ 1.1 & 14.58  $\pm$ 0.12 \\
     After  & -192.8 $\pm$ 1.1 & -193.7 $\pm$ 1.1 & 14.41  $\pm$ 0.12 \\
    \bottomrule
  \end{tabular}
\end{table}

\subsection{Energy calibration}
The energy calibration was performed using $X$ and $\gamma$-rays from an $^{241}$Am radioactive source~\cite{campbell86, toi}. The source was separated from the crystal by a PTFE sheet in between as shown in Fig.~\ref{f:setup} so that $\alpha$ particles from the source could be blocked. The digitizer was triggered when the PMT recorded a pulse above a certain threshold. The trigger rate was $\sim 6.3$~kHz when the threshold was set to 5 ADC counts above the pedestal level.

\begin{figure}[htbp]\centering
  \includegraphics[width=\linewidth]{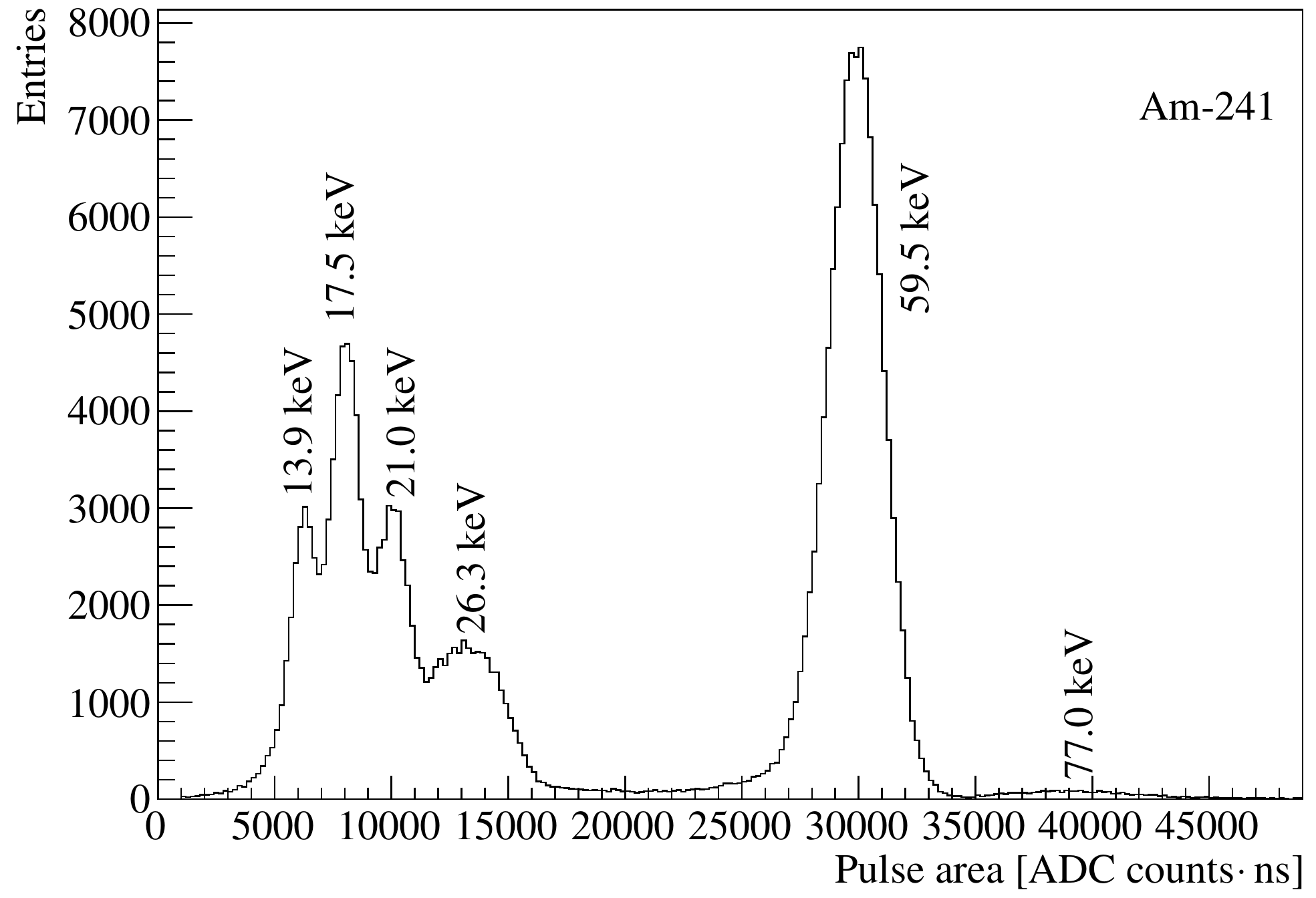}
  \caption{Energy spectrum of $^{241}$Am in the unit of ADC counts$\cdot$ns.}\label{f:spec}
\end{figure}

Each recorded waveform was 8008~ns long with a sampling rate of 1~GHz. About 1600~ns pre-traces were preserved before the rising edge of a pulse that triggered the digitizer so that there were enough samples before the pulse to calculate the averaged pedestal value of the waveform. After the pedestal was adjusted to zero the pulse was integrated until its tail fell back to zero. The integration had a unit of ADC counts$\cdot$ns.  The recorded energy spectrum in this unit is shown in Fig.~\ref{f:spec}.

The energy and origin of each peak were identified and summarized in Table~\ref{t:rPE}, based on Ref.~\cite{campbell86} and the \emph{Table of Radioactive Isotopes}~\cite{toi}.  The clear separation of the three $X$-rays peaks demonstrates a much better energy resolution than that of a NaI(Tl) detector working at room temperature.

\begin{table}[htbp]
  \caption{\label{t:rPE} Fitting results of $^{241}$Am peaks in the energy spectrum.}
  \begin{minipage}{\linewidth}\centering
  \begin{tabular}{cccccc}
    \toprule
    Type of & Energy & Mean & Sigma & FWHM \\
   radiation & (keVee) & (ADC$\cdot$ns) & (ADC$\cdot$ns) & (\%) \\
    \midrule
     \begin{tabular}{r} $X$-ray \\ $X$-ray \\ $X$-ray \\ $\gamma$-ray \\ $\gamma$-ray \\ Sum$^\ddagger$ \\
     \end{tabular} &
     \begin{tabular}{l} 13.9$^\dagger$ \\ 17.5$^\dagger$ \\ 21.0$^\dagger$ \\ 26.3 \\ 59.5 \\ 77.0 \\
     \end{tabular} &
     \begin{tabular}{r} 6303.6 \\ 8045.6 \\ 10076.0 \\ 13202.8 \\29817.6 \\39292.9 \\
     \end{tabular} &
     \begin{tabular}{r} 639.6 \\ 571.5 \\ 815.8 \\ 1598.5 \\1206.8 \\ 2674.9 \\
     \end{tabular} &
     \begin{tabular}{r} 23.9 \\ 16.7 \\ 19.1 \\ 28.5 \\ 9.5 \\ 15.9 \\
     \end{tabular} \\
     \bottomrule
     \end{tabular}
  \end{minipage}
  $^\dagger$ Intensity averaged mean of several $X$-rays near each other~\cite{campbell86, toi}.\\
  $^\ddagger$ Sum of $X$-rays and 59.5~keV $\gamma$-ray.
\end{table}

\begin{figure}[htbp]
  \includegraphics[width=\linewidth]{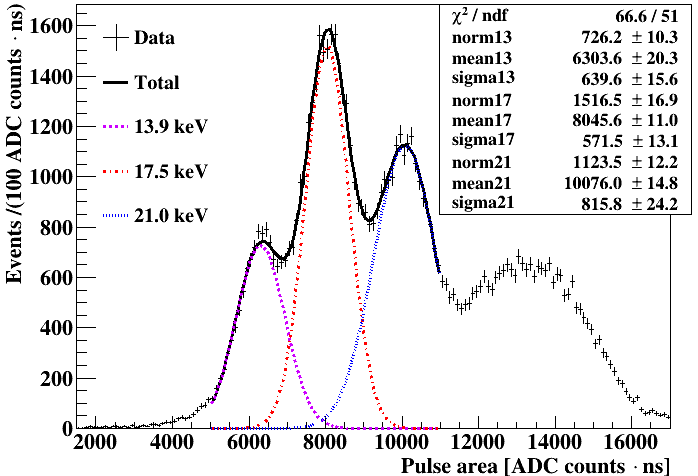}
  \includegraphics[width=\linewidth]{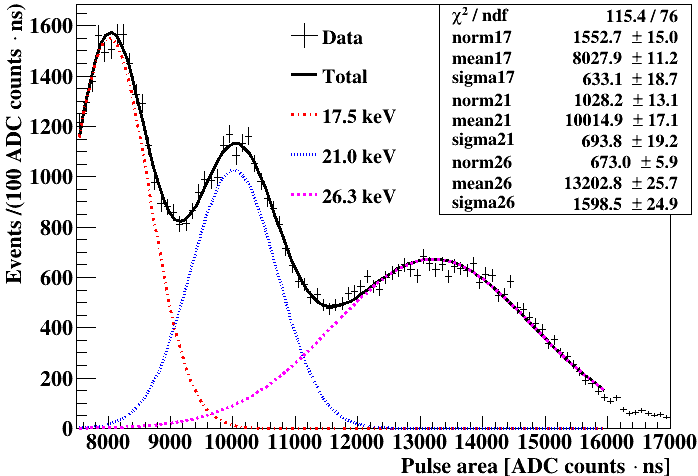}
  \includegraphics[width=0.98\linewidth]{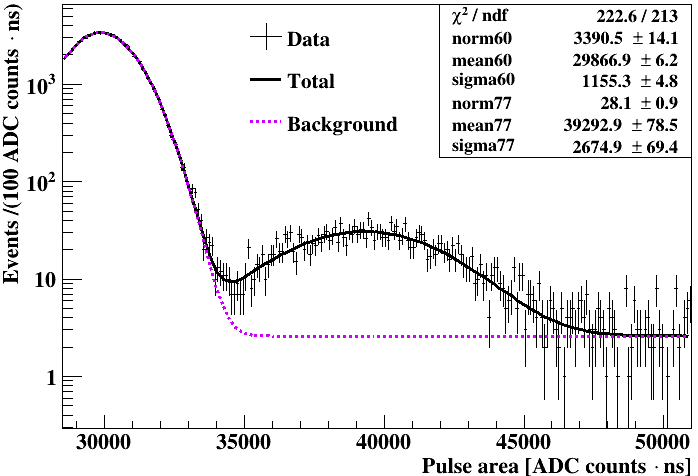}
  \includegraphics[width=0.98\linewidth]{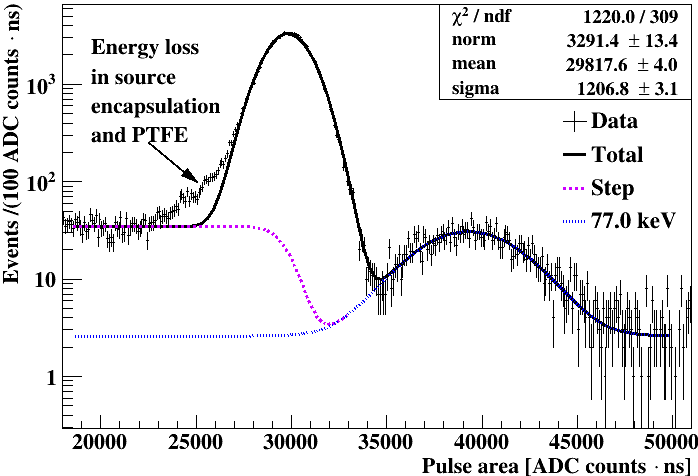}
  \caption{Fittings of individual $X$ or $\gamma$-ray peaks in the $^{241}$Am spectrum.}
  \label{f:fit}
\end{figure}

Peaks in Fig.~\ref{f:spec} were fitted with combinations of simple functions as shown in Fig.~\ref{f:fit} to extract their mean values and widths. The $X$-ray peaks at 13.8, 17.8 and 20.8~keV were fitted with three Gaussian distributions simultaneously (the first plot in Fig.~\ref{f:fit}), so were the 17.8, 20.8 and 26.3~keV peaks (the second plot in Fig.~\ref{f:fit}).  The 59.5 and 77.0~keV peaks were fitted with two Gaussian distributions on top of a horizontal line, the height of which was determined by the high energy side band of the 77.0~keV peak before the fitting. Part of the low energy side of the 59.5~keV peak was excluded from the fitting since it cannot be described by a pure Gaussian distribution (the third plot in Fig.~\ref{f:fit}). A Geant4-based Monte Carlo simulation~\cite{gears} revealed the origin of the tail on the low energy side of the 59.5~keV peak to be $\gamma$-rays that lost part of their energies in the source encapsulation and the PTFE plate in between the source and the crystal.

As shown in the last plot in Fig.~\ref{f:fit}, a different fitting method was tried for the 59.5~keV peak to verify the mean determined by the partial Gaussian fitting (the third plot in Fig.~\ref{f:fit}). Parameters of the function used to describe the 77.0 keV peak and its side bands were obtained from the third fitting and fixed in this fitting. The left side of the 59.5~keV peak was partially described by a step function associated with the Gaussian function~\cite{sag} used to fit the 59.5~keV peak:

\begin{equation}\label{e:sag}
  N_0 \erfc\left(\frac{x-\bar{x}}{\sigma}\right) + N_1 \exp(\frac{(x-\bar{x})^2}{2\sigma^2}),
\end{equation}
where the height of the step, $N_0$, was determined by the left side band of the 59.5~keV peak and fixed in the fitting. The normalization factor, $N_1$, the mean, $\bar{x}$ and the width, $\sigma$, of the Gaussian function were determined by the fitting. The difference of the means determined in these two methods is less than 0.2\%, the difference of the widths is less than 5\% The parameters obtained from the last fitting method was used for the later analysis.

\subsection{Light yield}
The fitted means and widths of the $X$ and $\gamma$-ray peaks in the $^{241}$Am spectrum in the unit of ADC counts$\cdot$ns were converted to the number of PE using the formula:
\begin{equation}
  \text{(number of PE)} = \text{(ADC counts} \cdot \text{ns)}/\bar{x},
  \label{e:m1pe}
\end{equation}
where $\bar{x}$ is the mean of the single-PE Gaussian distribution mentioned in Eq.~\ref{e:f1x}, also in the unit of ADC counts$\cdot$ns. The results are summarized in Table~\ref{t:rPE}.

\begin{figure}[htbp] \centering
  \includegraphics[width=\linewidth]{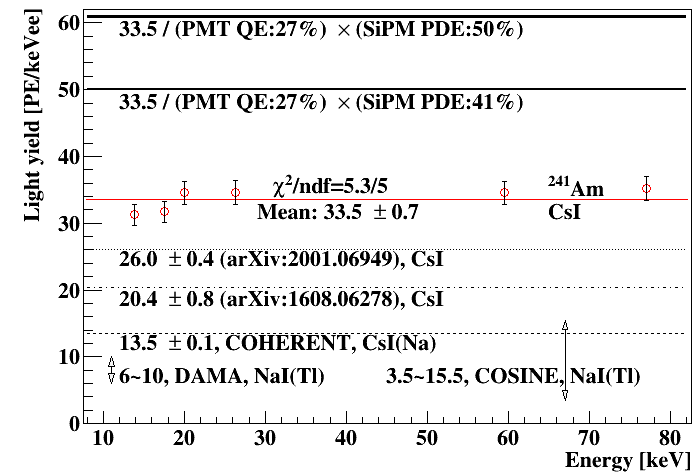}
  \caption{Currently and previously achieved~\cite{csi, csi20} light yields of undoped CsI at $\sim 77$~K together with the predicted ones with SiPM as light sensors. Those of the COHERENT CsI(Na)~\cite{coherent17}, DAMA/LIBRA NaI(Tl)~\cite{dama18} and COSINE NaI(Tl) detectors~\cite{cosine19} are plotted as well for comparison.}
  \label{f:LY}
\end{figure}

The light yield was calculated using the data in Table~\ref{t:rPE} and the following equation:
\begin{equation}
  \text{light yield [PE/keVee]} = \frac{\text{Mean [number of PE]}}{\text{Energy [keVee]}}.
  \label{e:ly}
\end{equation}
The obtained light yield at each energy point is shown in Fig.~\ref{f:LY}. The error bars are mainly due to the uncertainty of the mean value of the single-PE response used to convert the \textit{x}-axes of the energy spectra from ADC counts$\cdot$ns to the number of PE. The data points were fitted with a straight line to get an average light yield, which is 33.5 $\pm$ 0.7~PE/keVee.

\subsection{Non-linearity of light yield}
The non-linearity of both undoped CsI~\cite{Moszynski05} and NaI~\cite{Moszynski09} at 77~K have been investigated from 5.9~keV to 1.3~MeV with rather small crystals (a few mm in all dimensions). The results vary with crystals used in those studies. Some had less, others had more light yields at lower energies than that at 1.3~MeV. The difference ranges from 0 to 30\%.  As mentioned in Sec.~\ref{s:exp}, in an earlier measurement with the same crystal used in this study, a yield of $20.4\pm0.8$ PE/keVee was obtained in the energy range of [662, 2614]~keVee.  One of the purposes of this study was to verify the light yield of this larger crystal at a lower energy range. Thanks to the multiple low energy $X$-rays from the $^{241}$Am source, an even higher yield was achieved in the range of [13, 60]~keVee. The non-linearity observed so far seems not a concern for the application of undoped CsI at 77~K in neutrino and dark matter detections.

\section{Sensitivity of prototype detector to NSIs at the SNS}

Based on the measured light yield, the sensitivity to NSIs of a prototype detector made of $\sim10$~kg undoped CsI crystals placed at the SNS, ORNL, was estimated. General considerations of such a prototype have been discussed in a previous publication~\cite{csi20}. They will be briefly summarized here together with a detailed reasoning of adopting SiPMs instead of cryogenic PMTs as light sensors for the proposed prototype detector.

\subsection{Neutrino source}
The SNS is the world's premier neutron-scattering research facility. At its full beam power, about \num{1.5e14} \SI{1}{\GeV} protons bombard a liquid mercury target in \SI{600}{\ns} bursts at a rate of \SI{60}{\hertz}~\cite{coherent18}. Neutrons produced in spallation reactions in the mercury target are thermalized in cryogenic moderators surrounding the target and are delivered to neutron-scattering instruments in the SNS experiment hall.

As a byproduct, the SNS provides the world's most intense pulsed source of neutrinos peaked around a few of tens MeV~\cite{coherent18}. Interactions of the proton beam in the mercury target produce $\pi^+$ and $\pi^-$ in addition to neutrons. These pions quickly stop inside the dense mercury target. Most of $\pi^-$ are absorbed. In contrast, the subsequent $\pi^+$ decay-at-rest (DAR) produces neutrinos of three flavors. 

The sharp SNS beam timing structure ($\sim 1~\mu$s for prompt $\nu_\mu$, $\sim 10~\mu$s for delayed $\nu_e$ and $\bar{\nu}_\mu$~\cite{coherent18}) is highly beneficial for background rejection and precise characterization of those backgrounds not associated with the beam~\cite{Bolozdynya:2012xv}, such as those from radioactive impurities in a crystal.  Looking for beam-related signals only in the \SI{10}{\micro\s} window after a beam spill imposes a factor-of-2000 reduction in the steady-state background.

The COHERENT Collaboration occupies the ``Neutrino Alley'' located $\sim$20~m from the mercury target with contiguous intervening shielding materials and overburden eliminating almost all free-streaming pathways for fast neutrons which dominate beam-related backgrounds.  The prototype is assumed to be at the same location as the previous CsI(Na) detector.

\subsection{Crystal target}

\begin{figure}[htbp]\centering
  \includegraphics[width=0.5\linewidth]{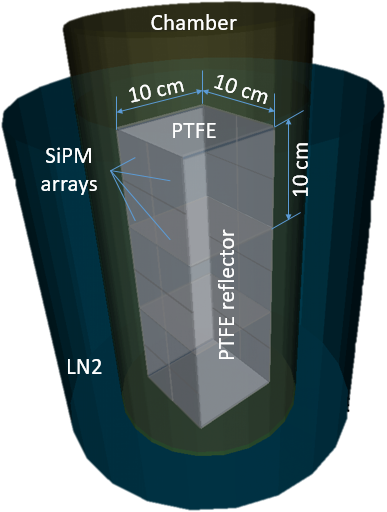}
  \caption{A simplified 3D drawing of a $\sim$10~kg detector module.}\label{f:g4m}
\end{figure}

About 10~kg undoped CsI operated at 77~K is assumed in this sensitivity analysis. Due to nearly identical scintillation mechanism and behavior (\cite{csi20} and references therein), undoped NaI can be another candidate. Multiple targets would be an even better choice as different isotopes in the targets help verify the neutron number dependence of the CEvNS cross section~\cite{coherent18}. The operation temperature is chosen for three reasons. The first is its convenience - LN2 cooling is conventional and economic. The second is to lower the DCR of SiPM arrays as a replacement of PMTs. This will be discussed in more detail in the following sections. The last is to utilize the high intrinsic light yields of undoped crystals at that temperature~\cite{Bonanomi52, Hahn53a, Hahn53, Sciver56, Beghian58, Sciver60, Fontana68, West70, Fontana70, Emkey76, persyk80, Woody90, Williams90, Wear96, Amsler02, Moszynski03, Moszynski03a, Moszynski05, Moszynski09, Sibczynski10, Sibczynski12, mikhailik15, csi, ess19}. The target mass is chosen to be similar to that of the COHERENT CsI(Na) detector~\cite{coherent17} for easy comparison.  Crystals in such a mass range can also be used as an optical module in a larger detector. Fig.~\ref{f:g4m} shows a simplified 3D drawing of such a module, where two opposite surfaces of a $10\times10\times10$ cm$^3$ crystal are covered by SiPM arrays, others are covered by PTFE light reflectors. The module with three cubic crystals can be directly submerged to LN2 or placed in a sealed chamber bathed in LN2.

\subsection{PMTs}
As liquid noble gas based dark matter detectors advance, there are quite some ultra-violet (UV) sensitive PMTs with a reasonable QE at 77~K available in the market, such as Hamamatsu R8778MODAY(AR)~\cite{yamashita10} and R11065, etc.~\cite{hotta14}.  Their performance in terms of light detection has been proven to be good enough in this and previous measurements~\cite{csi,csi20}. For example, a 1-PE trigger threshold of a detector can be translated to $\sim30$~eVee in energy, given a light yield of $33.5\pm0.7$~PE/keVee. This is much lower than the threshold of any existing inorganic scintillator based dark matter or neutrino experiment.

\begin{figure}[htbp]
  \includegraphics[width=\linewidth, trim=0 0 398 0, clip]{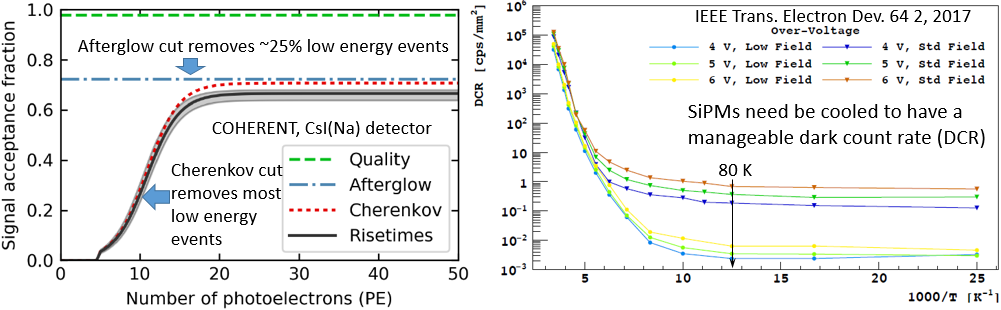}
  \caption{Detection efficiency of low energy events after each event selection criterion of the COHERENT CsI(Na) detector, adopted from Ref.~\cite{coherent17}.  Cherenkov cut removes most of the events near threshold.} \label{f:eff}
\end{figure}

However, energetic charged particles from natural radiation and cosmic rays can generate Cherenkov radiation when they pass through a PMT quartz (or fused silica) window. Given enough energy, a Cherenkov event can be easily distinguished from a scintillation event, since the former happens in a much shorter time window, the current pulse of which is much sharper than that of a scintillation event. However, close to the energy threshold, there are only a few detectable photons, which create a few single-PE pulses virtually identical in shape. The efficiency of pulse shape discrimination becomes lower and lower as the energy goes down. This is demonstrated clearly in Fig.~\ref{f:eff}, the detection efficiency of CEvNS events near the energy threshold of the COHERENT CsI(Na) detector, adopted from Ref.~\cite{coherent17}. The energy threshold of the detector was mainly limited by the Cherenkov event selection criterion instead of the light yield of the system.

In the COHERENT CsI(Na) detector, only one PMT was used. However, even if two PMTs are coupled to the two end surfaces of a cylindrical crystal, a request on coincident light detection in both of them cannot help remove Cherenkov events since the Cherenkov light created in one PMT can easily propagate to the other.

\subsection{SiPM arrays}
Two alternative sensors that do not generate Cherenkov radiation are avalanche photodiodes (APDs) and SiPMs. Made of silicon wafers, they can be much more radio-pure than PMTs, and do not need a thick SiO$_2$ window. APDs are a very attractive option~\cite{ess20} given their high PDE ($\sim$80\%). However, since they need to be operated in the linear mode, the gain is much less than those of PMTs and SiPMs, and cannot be triggered at single-PE level. On the other hand, a SiPM, which is basically an array of small APDs (micro cells) working in Geiger mode, is sensitive down to a single PE in each of its micro cell. The size of its micro cells has to be sufficiently small to avoid the situation when more than one photon hits the same micro cell. The space in between micro cells are not sensitive to photons. The peak PDE of a SiPM (up to 56\% at this moment~\cite{giovanetti17}) is hence smaller than that of an APD, but is typically higher than the peak QE of a PMT~\cite{jac14}.

Since covering a large area with a monolithic SiPM die is not possible mainly due to the production yield, a compromised solution is to tile several dies tightly together to form an array.   Given the same active area, a SiPM array uses less material, occupies less space, and can be made more radio-pure than a PMT. All make it a very attractive light sensor. Table~\ref{t:sipm} lists a few SiPM arrays that are already available in the market. All have an PDE that is higher than that of a PMT. Their gains are also very close to that of a typical PMT, which makes the signal readout much easier than that for an APD. More importantly, most of them have been tested in liquid argon or LN2 temperature (for example, Ref.~\cite{lc08, lightfoot09, rossi16, catalanotti15, johnson18} for SensL, Ref.~\cite{otono07, akiba09, igarashi16} for Hamamatsu, and Ref.~\cite{jj16} for KETEK SiPMs).  FBK SiPMs were proven working even down to 40~K with a good performance~\cite{aalseth17,acerbi17,giovanetti17}.  The light yield of the current system can be further improved by replacing PMTs (QE $\sim$ 27\%) with SiPM arrays (PDE: 40$\sim$50\%) to 50 or even 60 PE/keVee, shown as the top two lines in Fig.~\ref{f:LY}.

\begin{table}[htbp]\centering
  \caption{SiPM arrays available in the market possibly suitable for the proposed prototype detector}\label{t:sipm}
  \begin{tabular}{lcccc}\toprule
    SiPM & Microcell         & PDE$^\dagger$ & Largest array & Gain$^\diamond$ \\
    array& size ($\mu$m$^2$) &               & size (mm$^2$) & ($\times10^6$) \\\midrule
    S$^1$ J-series & 35$\times$35 & 50\% & $50.4\times50.4$ & 6.3\\
    S$^1$ C-series & 35$\times$35 & 40\% & $57.4\times57.4$ & 5.6\\
    H$^2$ S141xx &  50$\times$50 & 50\% & $25.8\times25.8$ & 4.7\\
    H$^2$ S133xx &  50$\times$50 & 40\% & $25.0\times25.0$ & 2.8\\
    K$^3$ PM3325 & 25$\times$25 & 43\% & $26.8\times26.8$ & 1.7 \\\bottomrule
  \end{tabular}
  \\\vspace{-0.2cm}\flushleft\hspace{0.2cm}$^\dagger$ @ $420\sim 450$~nm  \hspace{.4cm} $^\diamond$ @ 5 volt over-voltage
  \\\hspace{0.2cm}$^1$ SensL \hspace{1.7cm} $^2$ Hamamatsu \hspace{1.5cm} $^3$ KETEK
\end{table}

One major drawback of a SiPM array compared to a PMT is its high DCR at room temperature ($\sim$ hundred kHz). Fortunately, it drops quickly with temperature, and can be as low as 0.2~Hz/mm$^2$ below 77~K~\cite{aalseth17}, while the PDE does not change much over temperature~\cite{oto07, lc08, aki09, jan11}. However, a SiPM array that has an active area similar to a 3-in PMT would still have an about 100~Hz DCR at 77~K. A simple toy MC reveals that a 10-ns coincident window between two such arrays coupled to the same crystal results in a trigger rate of about $10^{-5}$~Hz. A further time coincidence with the SNS beam pulses would make the rate negligible.

Afterpulses~\cite{sipm} resulted from delayed releases of trapped electrons in metastable traps in a SiPM can mimic low energy events. But, just as DCR, they can be suppressed efficiently once coincident triggers are required.

Secondary photons with a wavelength range from 450~nm to 1600~nm can be emitted isotropically from a fired cell in a SiPM. Some of them can travel to a neighboring cell and cause optical crosstalks~\cite{oto07, lc08, aki09, jan11}.  All major manufacturers are actively improving their technologies to reduce the crosstalk rate, which ranges from 2\% to 27\% at this moment depending on the manufacturer, the size of micro cells, and the over-voltage applied. In general, smaller cell sizes and over-voltages cause less crosstalks but also smaller PDE.

However, the effect of optical crosstalk may be partially corrected for neutrino and dark matter induced low energy events close to the threshold, where the chance of one SiPM in an array to receive two photons at the same time is very low. By reading out individual SiPMs in an array, the pulse in a SiPM that is much larger than the pulse in any other SiPM has a high chance to be contaminated by crosstalk and can be regarded as a single PE. The effectiveness of this correction is verified by a simple optical simulation based on Geant4.

To summarize, there are three major technical advantages in the combination of cryogenic undoped crystals with SiPM arrays:
\begin{itemize}
  \item Cherenkov radiation from a PMT window is eliminated.
  \item Larger intrinsic light yields of cryogenic undoped crystals compared to those of doped ones at room temperature are utilized~\cite{csi20}.
  \item Larger PDE of SiPM arrays compared to QE of PMTs is utilized.
\end{itemize}
Meanwhile, the major drawbacks of SiPM arrays, such as DCR, afterpulses and optical crosstalks, can be kept under control with a reasonable amount of effort.

\subsection{Energy threshold}
Without Cherenkov radiation from PMTs, the energy threshold of the prototype detector is basically determined by its light yield. Assuming a conservative yield of 50~PE/keVee as shown in Fig.~\ref{f:LY}, and triggering on at least two photoelectrons in two different light sensors, the threshold can be roughly estimated as 2/50 = 40~eVee. To be more precise, a curve of trigger efficiency versus number of optical photons near the threshold was obtained with a toy Monte Carlo simulation detailed in Ref.~\cite{csi20}. Based on the curve, the trigger efficiency is about 50\% at 40~eVee.

\subsection{Quenching factor}
Another important detector property is the scintillation quenching factor for nuclear recoils. Since no systematic measurement of quenching factors for undoped crystals exists in such a low energy region, a constant quenching factor of 0.08 for NaI~\cite{xu15} and 0.05 for CsI ~\cite{coherent17} were taken from measurements with doped crystals, which can be translated to a 0.5~keV threshold for Na recoils, and a 0.8 keV threshold for Cs recoils. Given completely different scintillation mechanisms~\cite{csi20}, there is a possibility that scintillation quenching in undoped crystals is less serious than that in doped ones. For example, a very preliminary investigation~\cite{ess19} suggests a quenching factor of 0.1 for undoped CsI. The assumption here is hence conservative.

\subsection{Contributions of NSIs to CEvNS}
Assume that the NSIs between neutrinos and quarks in the target nuclei are mediated by a new vector boson, the differential CEvNS cross section for even-even nuclei can be parameterized from the Lagrangian in Eq.~\ref{e:nc} as~\cite{coherent17},
\begin{equation}\label{e:xs}
  \dv{\sigma}{T} = \frac{G^2_F M}{2\pi} Q^2_{w\alpha} F(q^2) \left[ 1 + \left( 1 - \frac{T}{E_\nu} \right)^2 - \frac{MT}{E^2_\nu} \right],
\end{equation}
where $M$ is the nuclear mass, $T$ is the nuclear recoil energy, $E_\nu$ is the neutrino energy, $F$ is the vector nuclear form factor assumed to be unity at the limit that the momentum transfer $q^2 \ll M^2$, and
\begin{equation*}\label{e:gv}
\begin{split}
   Q^2_{w\alpha} = [(g^p_V + 2\varepsilon^{u,V}_{\alpha\alpha} + \varepsilon^{d,V}_{\alpha\alpha}) Z
  + (g^n_V + \varepsilon^{u,V}_{\alpha\alpha} + 2\varepsilon^{d,V}_{\alpha\alpha})N] \\
  + \sum_{\beta\neq\alpha}[(2\varepsilon^{u,V}_{\alpha\beta}+\varepsilon^{d,V}_{\alpha\beta})Z + (\varepsilon^{u,V}_{\alpha\beta}+2\varepsilon^{d,V}_{\alpha\beta})N]^2
\end{split}
\end{equation*}
represents the weak charge for a neutrino of flavor, $\alpha$~\cite{coloma17prd}, where $Z$ and $N$ are proton and neutron numbers of the recoiled nucleus, $g^p_V=1/2-2\sin^2\theta_W$ and $g^n_V=-1/2$ are the SM NC vector couplings of neutrinos with protons and neutrons, respectively, $\theta_W$ being the weak mixing angle.

As an scattering experiment using neutrinos from pion decay at rest, the prototype has sensitivity only to NSI parameters $\varepsilon^{f,V}_{\alpha\beta}$ with $\alpha = \beta = e$ or $\mu$. As an example, we consider two scenarios: varying $\varepsilon^{u,V}_{ee}$ and $\varepsilon^{d,V}_{ee}$ with all other couplings assumed to be zero, and varying $\varepsilon^{u,V}_{ee}$ and $\varepsilon^{u,V}_{\mu\mu}$ with all other couplings assumed to be zero. The former scenario considers variations in the two least-well measured parameters while with the later non-zero couplings could resolve current tension in neutrino oscillation experiments~\cite{coloma17prd}.

Note that the standard three-flavor model of neutrino mixing is assumed, and that the baseline is too short for significant flavor transition.

From Eq.~\ref{e:xs}, one can see that the presence of non-zero NSI results in an overall scaling of the event rate, either enhancement or suppression, rather than a spectral distortion.

\subsection{Sensitivity to NSIs}

\begin{figure}[htbp]
  \includegraphics[width=\linewidth]{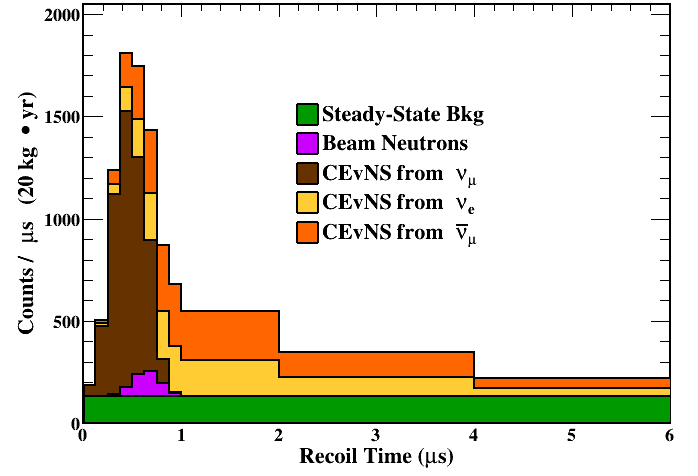}
  \includegraphics[width=\linewidth]{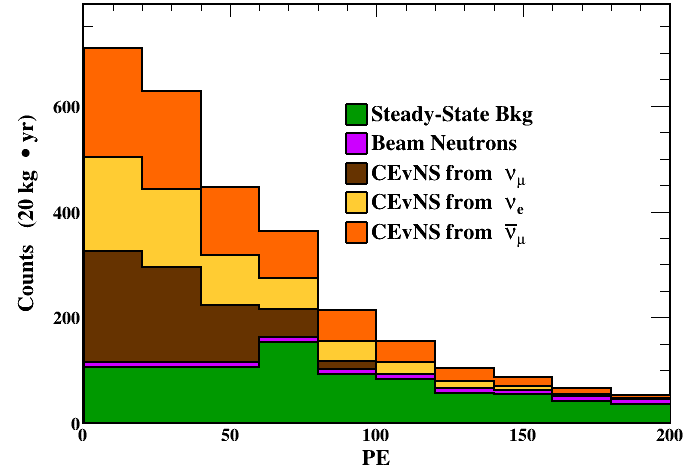}
  \caption{The predicted event samples of a 10 kg CsI crystal after two years at the SNS. The CEvNS counts from different components of the flux are shown in brown, $\nu_\mu$, gold, $\nu_e$, and orange, $\bar{\nu}_\mu$. Due to the timing structure of the beam, the CEvNS from $\nu_\mu$ and $\nu_e$ scatters are readily separated allowing for detailed study of NSI parameters.}\label{f:PredInT}
\end{figure}

We perform a spectral fit in both the recoil energy and time domains simultaneously to the estimated event rate distribution, shown in Fig. \ref{f:PredInT} as a function of the recoil time and PE. As the time distribution for CEvNS from $\nu_e$ is characteristically different from that for CEvNS from $\nu_\mu$ and $\bar{\nu}_\mu$ at the SNS, such a fit can distinguish between NSI effects caused by non-zero $\epsilon^{f,V}_{ee}$ and $\epsilon^{f,V}_{\mu\mu}$.

This fit maximizes a log-likelihood while profiling over systematic uncertainties. We assume a 10$\%$ uncertainty on the neutrino flux, a $5\%$ uncertainty on the CEvNS rate due to nuclear form factor uncertainty, and a 25$\%$ uncertainty on the normalization of the beam-related neutron background~\cite{coherent17}. Additionally, a 25$\%$ uncertainty on the quenching factor is assumed. Though, with such a low threshold, this variation in quenching factors only adjusts the CEvNS rate by 3.5$\%$.

\begin{figure}[htbp]\centering
  \includegraphics[width=0.8\linewidth]{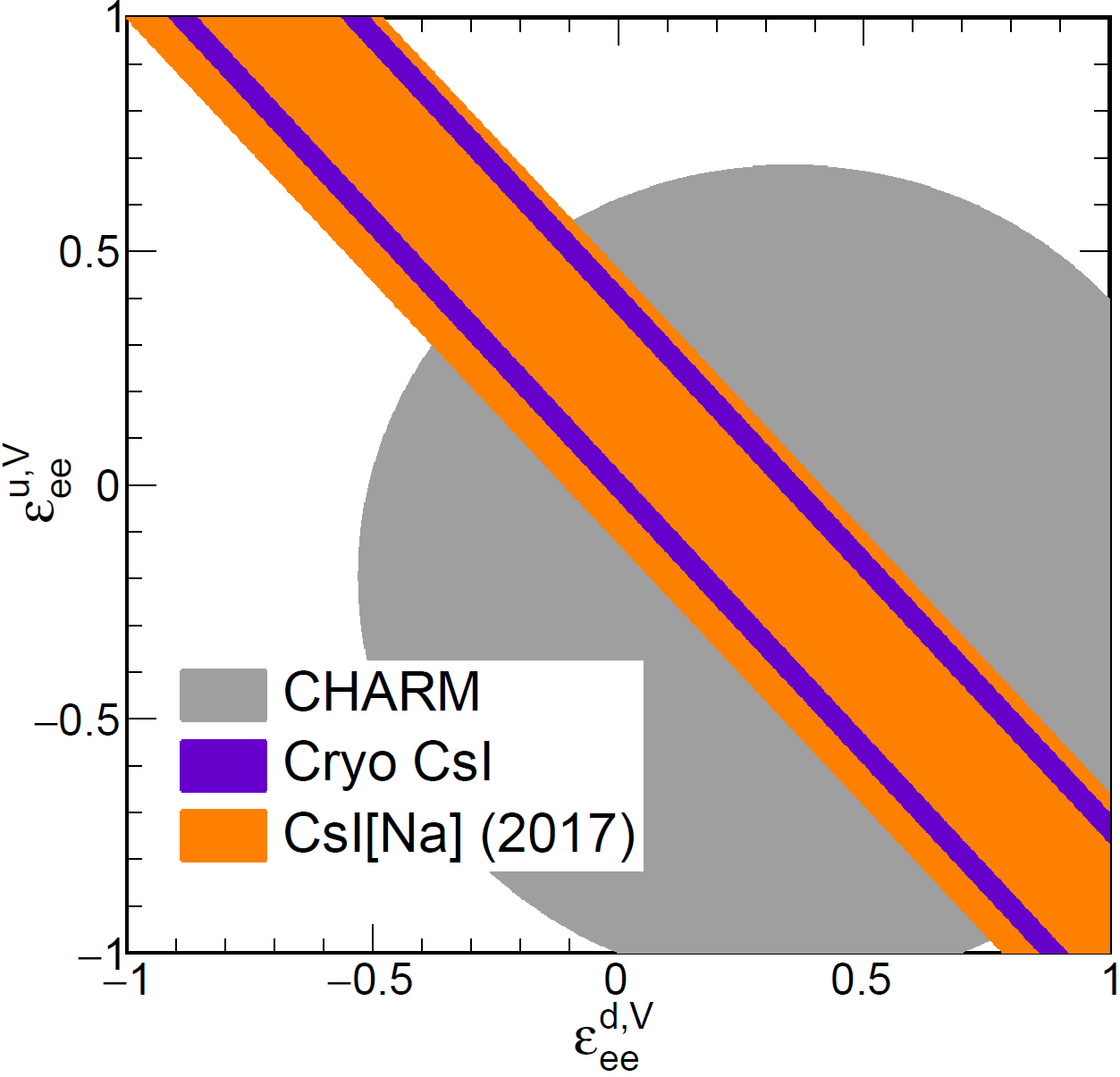}
  \caption{Constraints on two NSI parameters, $\varepsilon_{ee}^{u,V}$ and $\varepsilon_{ee}^{d,V}$ from the proposed detector compared to the ones from the CHARM experiment and the COHERENT CsI(Na) detector.}\label{f:edv}
\end{figure}

Our estimated sensitivity to ($\varepsilon^{u,V}_{ee},\varepsilon^{d,V}_{ee}$), after two years of a 10 kg detector at the SNS at 90$\%$ confidence, is shown in Fig.~\ref{f:edv} along with current constraints. The detector would give constraints much tighter than the initial COHERENT result.

\begin{figure}[htbp]
  \includegraphics[width=\linewidth]{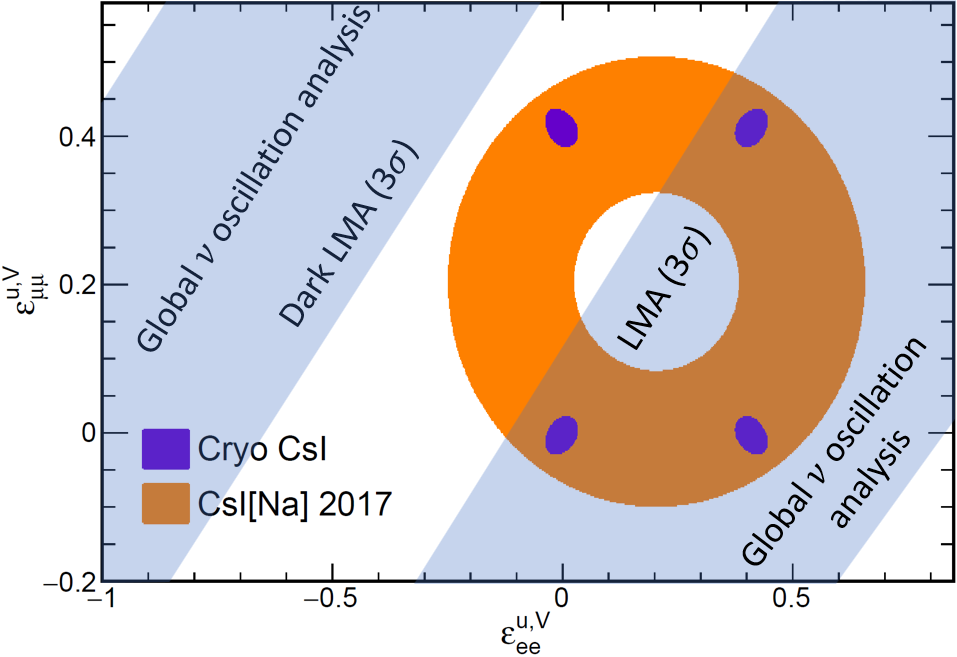}
  \caption{Constraints on two NSI parameters, $\varepsilon_{\mu\mu}^{u,V}$ and $\varepsilon_{ee}^{u,V}$ from the proposed detector compared to the one from the COHERENT CsI(Na) detector. Both the normal and dark LMA solutions from a global neutrino oscillation analysis are overlaid to demonstrate the effectiveness of COHERENT detectors in resolving degeneracy introduced by NSIs.}\label{f:euv}
\end{figure}

The existence of such non-standard interactions of neutrinos could confound the interpretation of data from neutrino oscillation experiments. The constraint estimated in $(\varepsilon^{u,V}_{ee},\varepsilon^{u,V}_{\mu\mu})$ parameter space is shown in Fig.~\ref{f:euv}. Oscillation experiments alone can not discern between the accepted measured $\theta_{12}$ in the first octant, and the ``Dark LMA" solution, where observed data would imply a $\theta_{12}$ in the second octant along with non-zero values of specified NSI couplings~\cite{coloma17, coloma17prd}. Also shown is the parameter space consistent with the Dark LMA solution in comparison to our predicted sensitivity. The constraint improves separation between parameter space allowed by CEvNS scattering experiments and that consistent with the Dark LMA solution further cementing the value measured in the LMA assumption. 


\section{Conclusion}
The sensitivity of a 10~kg prototype detector based on cryogenic inorganic scintillating crystals coupled to SiPM arrays to probe NSIs through CEvNS detection at the SNS, ORNL, was investigated. After two years of data taking, the presumed detector can pose much more stringent constraints on the least constrained NSI parameters than the existing ones from the COHERENT CsI(Na) detector. The constraints can also be used to break the degeneracy in neutrino oscillation parameters in the framework of NSIs, hence help in interpolating neutrino oscillation data in general. The key technical advantages of the prototype include much higher light yields of undoped crystals at 77~K compared to those of doped ones at room temperature, and a complete elimination of Cherenkov radiation originated from PMTs that seriously limits the energy threshold of current inorganic scintillator detectors. As one of the initial steps to verify the feasibility of the proposed technique, the light yield of an undoped CsI crystal at about 77~K was measured to be $33.5\pm0.7$~PE/keVee in the energy range around [13, 60]~keVee.  This was about three times higher than those achieved with the COHERENT CsI(Na) detector~\cite{coherent17}, and the DAMA~\cite{dama18} and COSINE~\cite{cosine19} NaI(Tl) detectors.

\begin{acknowledgements}
  This work is supported by the National Science Foundation (NSF), USA, award PHY-1506036, and the Grant-in-Aid for Encouragement of Young Scientists (B), No. 26800122, MEXT, Japan. Computations supporting this project were performed on High Performance Computing systems at the University of South Dakota, funded by NSF award OAC-1626516.
\end{acknowledgements}

\bibliographystyle{spphys} 
\bibliography{ref}

\end{document}